\documentstyle[twoside,fleqn,espcrc2]{article}

\input epsf   

\newcommand{\aar}{${}^{\footnotesize a}$}
\newcommand{\la}{\left\langle}
\newcommand{\ra}{\right\rangle}

\newcommand{\psihat}{\widehat{\psi}}
\newcommand{\psibar}{\overline\psi}
\newcommand{\psihatbar}{\widehat{\psibar}}

\newcommand{\ovtitle}{
\vskip -20 mm
\noindent {\bf hep-lat/9311062} \par
\noindent {\bf DFTUZ 93/18} \par
\noindent November 1993 \par
\vskip 5 mm
}

\title{
\ovtitle
Phase diagram of a lattice $SU(2) \times SU(2)$ scalar-fermion model
using the Zaragoza fermions\thanks{Talk presented
by F. Lesmes in Lattice'93.}}

\newcommand{\Zaragoza}{Departamento de F\'\i sica Te\'orica,
		       Universidad de Zaragoza,\\
		       50009 Zaragoza, Spain}
\newcommand{\Paris}{Laboratoire de Physique Th\'eorique et Hautes Energies,\\
		       Universit\'e de Paris XI, 91405 Orsay Cedex, France}

\author{J.L. Alonso\address\Zaragoza,
	Ph. Boucaud\address\Paris,
	F. Lesmes\aar\  and
	E. Rivas\aar}

\begin{document}

\begin{abstract}
We present a calculation of the phase diagram of
a $SU(2) \times SU(2)$ chiral Yukawa model with massless
decoupled doublers, using a saddle point approach, both for small and large
Yukawa coupling.
Some preliminary MonteCarlo results are also shown.
\end{abstract}

\maketitle

\section{INTRODUCTION}

A lattice regularization of the Standard Model (SM)
is necessary not only to establish the existence of chiral gauge theories
outside perturbation theory, but also to understand various properties of the
fermion mass generation by spontaneous symmetry breaking such as,
for instance, the
upper bounds on top and Higgs masses\cite{Pet93}.
For this purpose it is sufficient to study the fermion-scalar
sector of the SM which is essentially a Chiral Yukawa Model (CYM).

In this talk, we determine the phase structure of a $SU(2) \times SU(2)$
CYM, using Zaragoza fermions\cite{Alo91}.
We decided to freeze the radial mode of the scalar
field. This corresponds to the choice of an infinite bare quartic coupling.
For a comparison with the phase structure of other related models
see\cite{Pet93,Bock90}.

\section{LATTICE REGULARIZED CHIRAL YUKAWA MODEL}

Consider $N_d$ $SU(2)$ fermion doublets $\psi$, and a $2\times 2$ $SU(2)$
matrix, $\phi$.
 The action can be cast into the form \cite{Alo91}
\begin{equation}
S(\psi,\phi) = S_B(\phi) + S_F(\psi) + S_Y(\psihat,\phi),
\end{equation}
where
\begin{equation}
S_B(\phi) = - {\kappa \over 2}\sum_{x,\mu}
Tr \left( \phi_{x+\hat\mu}^\dagger \phi_x
+ \phi_x^\dagger \phi_{x+\hat\mu} \right)
\end{equation}
is the lattice action for a scalar with frozen radial modes.
$ S_F(\psi)$, the free fermionic action is
\begin{equation}
S_F(\psi) = {1\over 2} \sum_{\psi,x,\mu} \left(
\psibar_x\gamma_\mu \psi_{x+\hat\mu} -
\psibar_{x+\hat\mu}\gamma_\mu\psi_{x}
\right),
\end{equation}
and the interaction term is
\begin{equation}
S_Y(\psihat,\phi) = y \sum_{\psi,x}\left(\psihatbar_{Lx}\phi_x\psihat_{Rx}
 		+ \psihatbar_{Rx}\phi_x^\dagger\psihat_{Lx}\right).
\end{equation}
Our way of implementing the decoupling of doublers is based on the use of the
$\psihat$ fields. In momentum space,
\begin{equation}
\psihat(k) = F(k)\psi(k), \quad
F(k) = \prod_\mu \cos\left( {k_\mu \over 2} \right).
\end{equation}
In coordinate space this corresponds to an average of fields $\psi$ over
an elementary hypercube
\begin{equation}
\psihat_x = {1 \over 2^4} \sum_b \psi_{x+b},\quad b_\mu = 0,1.
\end{equation}

The model has a global chiral $SU(2) \times SU(2)$ symmetry
\begin{eqnarray}
& & \psi_{L}(x) \to \Omega_L \psi_{L}(x), \qquad
\psi_{R}(x) \to \Omega_R \psi_{R}(x), \nonumber \\
& & \phi(x) \to \Omega_L \phi(x) \Omega_R^\dagger,
\end{eqnarray}
where $\Omega_R,\Omega_L \in SU(2)$. (A corresponding transformation
acts on the barred fermions).

\section{DECOUPLING OF DOUBLERS}
The model has been constructed to show decoupling at tree level.
The fermion-scalar vertex  is given by
\begin{equation}
V(k,k') = y F(k) F(k'),\label{vertex}
\end{equation}
and the particular choice of $F(k)$ makes it vanish
for every doubler fermion (any $k_{\mu}$ or $k_\mu' =
\pi$).
It can also be shown that at any order in perturbation theory,
the renormalized Green functions are the same as the continuum ones.
The proof\cite{Alo91} is based on the Riesz power counting theorem.
There are also ($2^4 - 1$) Golterman-Petcher like
symmetries\cite{Alo91,Gol89}
which can be used to argue that the doubler fermions also
decouple in the nonperturbative regime, provided a sensible continuum limit
exists.

An important characteristic of our decoupling method is that the
$U$, $S$ and $\Delta \rho$ parameters
(which describe the radiative corrections to
the electroweak interaction observables coming
from the Higgs sector)
do not receive, at one loop level,
contributions from the counterterms needed to recover
gauge invariance\cite{Alo93}.
This allows us, using only the scalar-fermion sector of the SM, to
compute $\Delta \rho$ non-perturbatively.
Prior to do this and the computation of upper bounds on top and Higgs masses
a phase diagram is required.

\section{ MEAN FIELD CALCULATION OF THE PHASE DIAGRAM}

We have performed an analytic calculation of the phase diagram,
using mean field techniques, both for small and large values of the Yukawa
coupling $y$.

There are two essential steps.

\noindent $\bullet$   First. With the aid of two auxiliary scalar
fields,
one performs the integration over the scalar field $\phi$.

$\bullet$$\bullet$ For small $y$, one expands
$\exp\left\{-S_Y\right\}$
arround 1, up to four fermion terms before integrating out the scalar field.

$\bullet$$\bullet$ For large $y$, auxiliary fermions $\eta$ are used in
order to  change the
coupling parameter to $1/y$.
In particular, for one fermion doublet,
\begin{eqnarray*}
\exp\left\{-S_Y\right\} \propto \int \left [ d  \eta  d\overline\eta  \right ]
exp \left\{ \overline\eta\psihat+\psihatbar\eta\right\}
\end{eqnarray*}
\begin{equation}
{} \times \exp \left\{  {1\over y}\overline\eta \left( \phi^+P_R+
\phi P_L\right)\eta \right\}.
\end{equation}
One can now perform an expansion of the exponential around 1, up to four
fermion
terms, then integrate out the scalar field.
Finally one integrates over the auxiliary $\eta$.

\noindent $\bullet$ Second.  After replacing the auxiliary fields
with their saddle point solutions
one is able to calculate (in both regimes) the fermion determinant
and therefore the Free energy.

\begin{figure}[tbh]
\centerline{
\epsfxsize=7.4 cm
\epsfbox[72 94 521 361]{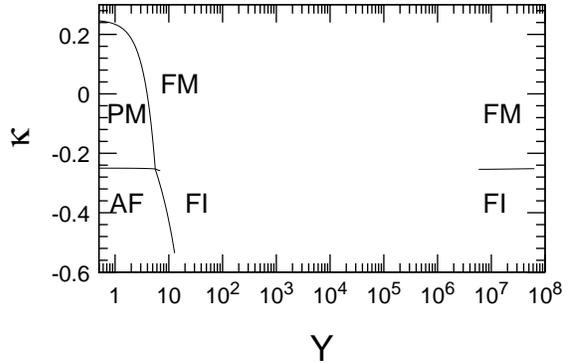}
}
\caption{Mean field calculation of the phase diagram.
The number of fermion doublets ($N_d$) is two.}
\label{full}
\end{figure}

The order parameters we have used to distinguish the phases are
\begin{equation}
\la\phi\ra = \la{1\over V}\sum_n\phi_n\ra
\end{equation}
and
\begin{equation}
\la\phi_{st}\ra = \la{1\over V}\sum_n\epsilon_n\phi_n\ra,
\end{equation}
where $\epsilon_n = {(-1)}^{n_1 + \dots + n_4}$.

There are four phases:

\noindent $\bullet$ FM: In this phase $\la \phi \ra \not= 0$ and $\la \phi_{st}
\ra = 0$.

\noindent $\bullet$ PM: Here, $\la \phi \ra = \la \phi_{st} \ra = 0$.

\noindent$\bullet$  AF: This phase is characterized by
              $\la \phi \ra = 0$ and $\la \phi_{st} \ra \not= 0$.

\noindent $\bullet$ FI: Both parameters are different from zero.

\begin{figure}[tbh]
\centerline{
\epsfxsize=7.4 cm
\epsfbox[72 94 582 442]{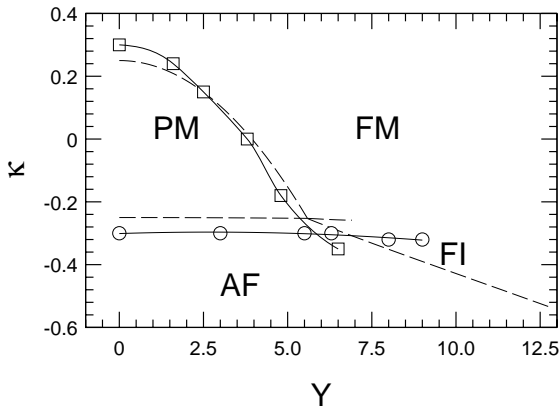}
}
\caption{Phase diagram. Dashed lines correspond to the mean field calculation.
Square and circle symbols are numerical results in a $4^4$ lattice. $N_d = 2$.
}
\label{numerical}
\end{figure}

The mean field results are presented
in figures \ref{full} and \ref{numerical},
where the range of validity for
both small and large $y$ expansions can be appreciated
($y<11$ and $y > 6 \times 10^6$ respectively).
For small $y$  the expected FM--PM--AFM structure appears together
with a quadruple point and a FI phase.
For large $y$ only two phases appear,
FM and FI, so that the shape of funnel is excluded for this model.

The phase transitions are second order within this mean field approach.
In the same approach, we can understand why the PM-AF transition line is
almost flat. This is a consequence of the decoupling.
In fact, in the AF phase, the scalar field takes the form $\phi_n =
v_{st}\epsilon_n$,
thus it always carries momentum $\pi$.
Due to momentum conservation, when this scalar interacts with two fermions,
the latter  must  be either a  physical fermion and a doubler or two different
doublers.
In either case, the interaction is supressed as doublers are decoupled.
So, the critical hopping
parameter $\kappa$ is not affected by the Yukawa coupling $y$.
This behaviour is confirmed by the MonteCarlo results
(Figure \ref{numerical}).

\section{MONTECARLO SIMULATION}

The mean field calculation gives us a feeling for the phase diagram, but
in order to have quantitative results we are doing a numerical simulation
of this model with two doublets of fermions, using a Hybrid-MonteCarlo
algorithm.
Figure \ref{numerical} shows some preliminary results for a
$4^4$ lattice. They are in good agreement with the mean field computations.
Current work includes  a MonteCarlo simulation in a $8^4$ lattice.
Using this larger lattice we expect first to get a clearer
determination of the phase transition lines, second
to calculate upper bounds on top and Higgs masses,
and third to study non-perturbatively the $\Delta \rho$ parameter.

\end{document}